\documentclass[prb,twocolumn,longbibliography]{revtex4-1}

\usepackage{braket}
\usepackage{mathtools}
\usepackage[dvipsnames]{xcolor}

\usepackage{xargs}                      
\usepackage{amsmath}
\usepackage{amssymb}
\usepackage{bbold}
\usepackage{comment}

\newcommand{\beq}{\begin{equation}}
\newcommand{\eeq}{\end{equation}}
\newcommand{\beqa}{\begin{eqnarray}}
\newcommand{\eeqa}{\end{eqnarray}}
\newcommand{\ben}{\begin{enumerate}}
\newcommand{\een}{\end{enumerate}}
\newcommand{\bfg}{\begin{figure}}
\newcommand{\efg}{\end{figure}}
\newcommand{\Fig}[1]{Fig.~\ref{fig:#1}}
\newcommand{\Eq}[1]{Eq.~\ref{eq:#1}}

\newcommand{\avg}[1]{\langle #1 \rangle}
\newcommand{\Tr}[1]{\text{Tr}\big[#1\big]}
\newcommand{\partZ}{\mathcal{Z}}

\usepackage{graphicx}

\begin{document}

\title{Hybrid Purification and Sampling Approach for Thermal Quantum Systems}

\author{Jing Chen}
\affiliation{Center for Computational Quantum Physics, Flatiron Institute, New York, NY 10010 USA}

\author{E.M.\ Stoudenmire}
\affiliation{Center for Computational Quantum Physics, Flatiron Institute, New York, NY 10010 USA}

\date{\today}

\begin{abstract}
We propose an algorithm which combines the beneficial aspects of two different methods for studying 
finite-temperature quantum systems with tensor networks. 
One approach is the ancilla method, which gives high-precision results but
scales poorly at low temperatures. The other method is the minimally
entangled typical thermal state (METTS) sampling algorithm which scales better
than the ancilla method at low temperatures and can be parallelized, 
but requires many samples to converge to a precise result.
Our proposed hybrid of these two methods purifies physical sites in a small central spatial region 
with partner ancilla sites, sampling the remaining sites using the METTS algorithm.
Observables measured within the purified cluster have much lower sample variance than in the
METTS approach, while sampling the sites outside the cluster reduces their entanglement and 
the computational cost of the algorithm. The sampling steps of the algorithm remain straightforwardly 
parallelizable. 
The hybrid approach also solves an important technical issue with METTS  
that makes it difficult to benefit from quantum number conservation.
By studying \mbox{$S=1$} Heisenberg ladder systems, 
we find the hybrid method converges more quickly than
both the ancilla and METTS algorithms at intermediate temperatures 
and for systems with higher entanglement.

\end{abstract}

\maketitle

Tensor networks are an approach for many-body quantum systems 
whose effectiveness is usually associated with low-energy states, which
are known to have limited entanglement.
Yet by using wavefunctions time-evolved from product states, 
one can apply tensor networks to phenomena involving wide energy ranges,
including out-of-equilibrium quantum systems and equilibrium systems at temperature $T$, 
which is the focus of this work.
By using tensor networks for finite temperature systems,
 one can straightforwardly handle models with itinerant fermions or 
frustrated spin interactions, for which
quantum Monte Carlo is often severely limited due to the sign problem.
The tradeoff is that tensor network methods are presently limited to 
one- or two-dimensional quantum systems, with the two-dimensional case
requiring significant effort.
It therefore becomes desirable to reduce the cost of tensor network methods
enough that large two-dimensional systems can be studied.

The two leading approaches for studying finite-temperature systems with
tensor networks are the ancilla method (also known as the purification method),
\cite{Zwolak:2004,Verstraete:2004d,Feiguin:2005}
and the minimally entangled typical thermal states (METTS) sampling
algorithm.\cite{White:2009,Stoudenmire:2010}
Both methods are based on well-established tensor network techniques 
for evolving quantum states in imaginary time.

\begin{figure} 
\includegraphics[width=\linewidth]{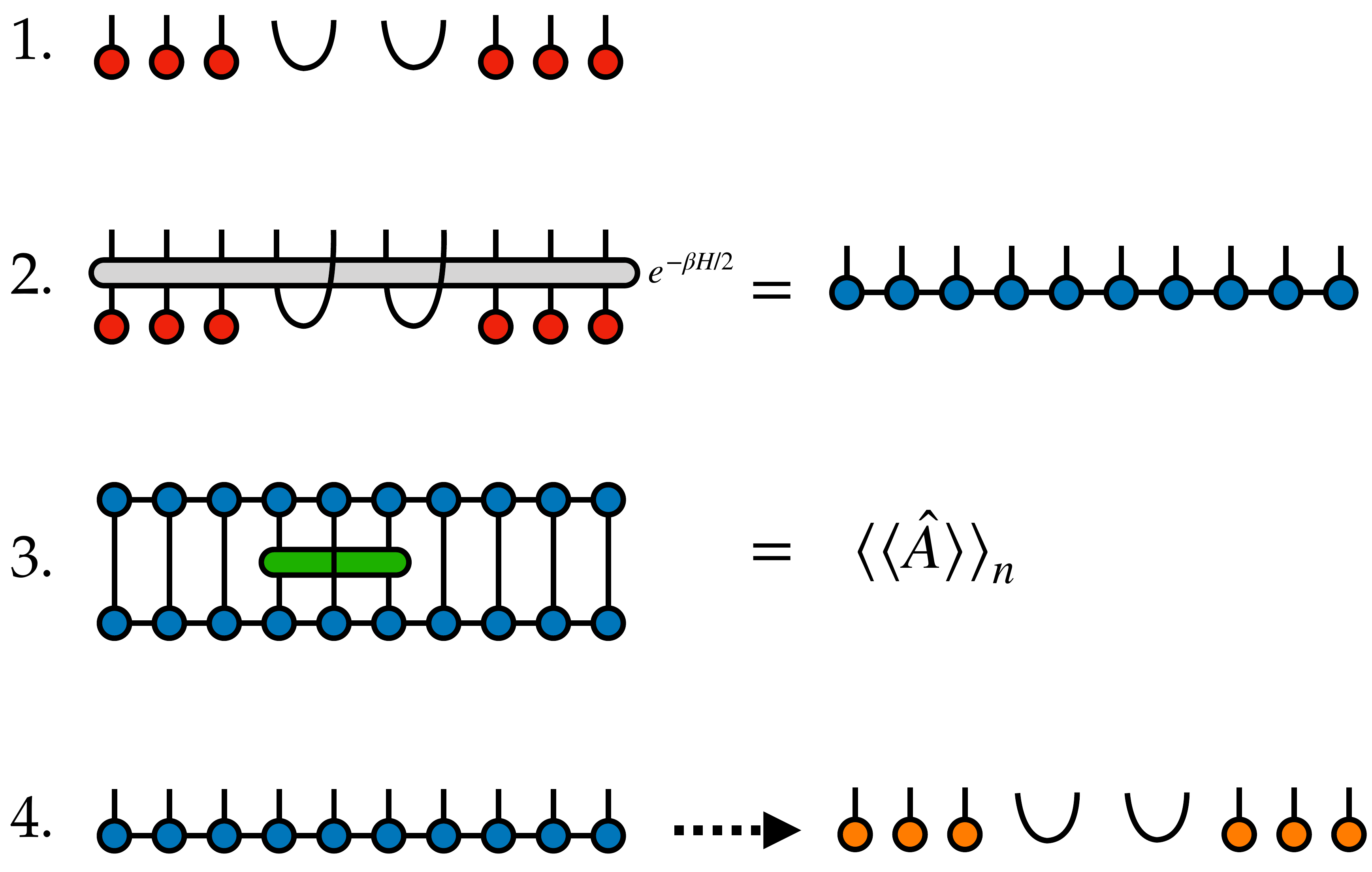}
\caption{Illustration of the steps of the hybrid purification-sampling method.
Step 1 begins with a pure state where the center $2W$ ``cluster sites'' are 
in maximally entangled pairs and the remaining ``environment sites'' are in a product state. 
In step 2, the state is evolved for an imaginary time $\beta/2$, with the evolution
operator acting only on the physical sites. 
Step 3 involves the measurement of observables on physical sites of the cluster.
In step 4, one collapses the environment sites back into product states through
projective measurements, 
then replaces the cluster sites with maximally entangled pairs.}
\label{fig:illustration}
\end{figure} 

In the ancilla method, one evolves the many-body identity operator for an
imaginary time of $\beta/2$, resulting in an approximation of the operator $e^{-\beta H/2}$.
Two copies of this operator are then traced with local observables to compute 
thermal expectation values.
The method is typically formulated by viewing the initial 
identity operator on $N$ sites as a pure state of maximally entangled pairs on $2N$ sites, 
with every other site viewed as an artificial, ancilla site whose role is to thermalize the 
physical sites. Though the approach works
very well at higher temperatures, at low temperatures the cost becomes
extremely high compared to ground-state techniques.

In the METTS algorithm, one evolves a product state wavefunction for an imaginary
time of $\beta/2$, resulting in an entangled state called a METTS.\cite{White:2009} 
Then a projective measurement of each lattice site is used to sample a new
product state, which is evolved to produce the next METTS. The algorithm
is therefore a type of quantum Monte Carlo where each sample is an entangled state. 
When implemented using matrix product state (MPS) 
techniques, the METTS algorithm scales similarly to the density matrix renormalization
group (DMRG) at low temperatures, making it much more efficient than the
ancilla method at low enough temperatures. 
However, the additional sampling overhead of the METTS
algorithm makes this crossover temperature very low.

It is conceivable that the ancilla and METTS techniques could be 
combined to realize the best aspects of both algorithms, since both involve imaginary 
time evolution of specially chosen states with tensor network methods.  
Indeed, one of the original motivations of METTS was sampling from the wavefunction produced by the ancilla method.\cite{White:2009}

In what follows, we will hybridize of the ancilla and METTS approaches by dividing the system spatially
into a small cluster embedded within the rest of the system (the ``environment''). The degrees of freedom in the cluster are purified initially as product of maximally entangled states, while the degrees of freedom in the environment are sampled over product states similar as METTS, so that the entanglement is reduced within the environment and also between cluster and environment. Due to the purification within the cluster, local 
observables measured there converge much more quickly with the number of states sampled 
compared to the METTS approach. 
Meanwhile, degrees of freedom in the environment are less entangled than in the ancilla approach,
saving valuable computational resources.
This hybrid method strikes a balance between the number of samples needed and the
cost of producing each sample, which  can be controlled 
by tuning the cluster size. 
When the ``system cluster'' shrinks to zero spatial size, 
the method reduces to the METTS algorithm. 
In the other limit of the cluster covering the entire system, 
the method becomes the ancilla method.

\section{Hybrid Purification-Sampling Method}

To study the physics of finite-temperature quantum systems, we will work in the canonical
ensemble where the central quantity is the finite-temperature density matrix
\begin{align}
\rho_\beta = \frac{1}{\partZ} e^{-\beta H} \ .
\end{align}
Here $H$ is the Hamiltonian operator governing our system and $\beta=1/T$. 
The partition function $\partZ$ is
defined as \mbox{$\partZ=\text{Tr}[e^{-\beta H}]$} to ensure $\rho_\beta$ has unit trace.

\begin{figure}[t] 
\centering
\includegraphics[width=\linewidth]{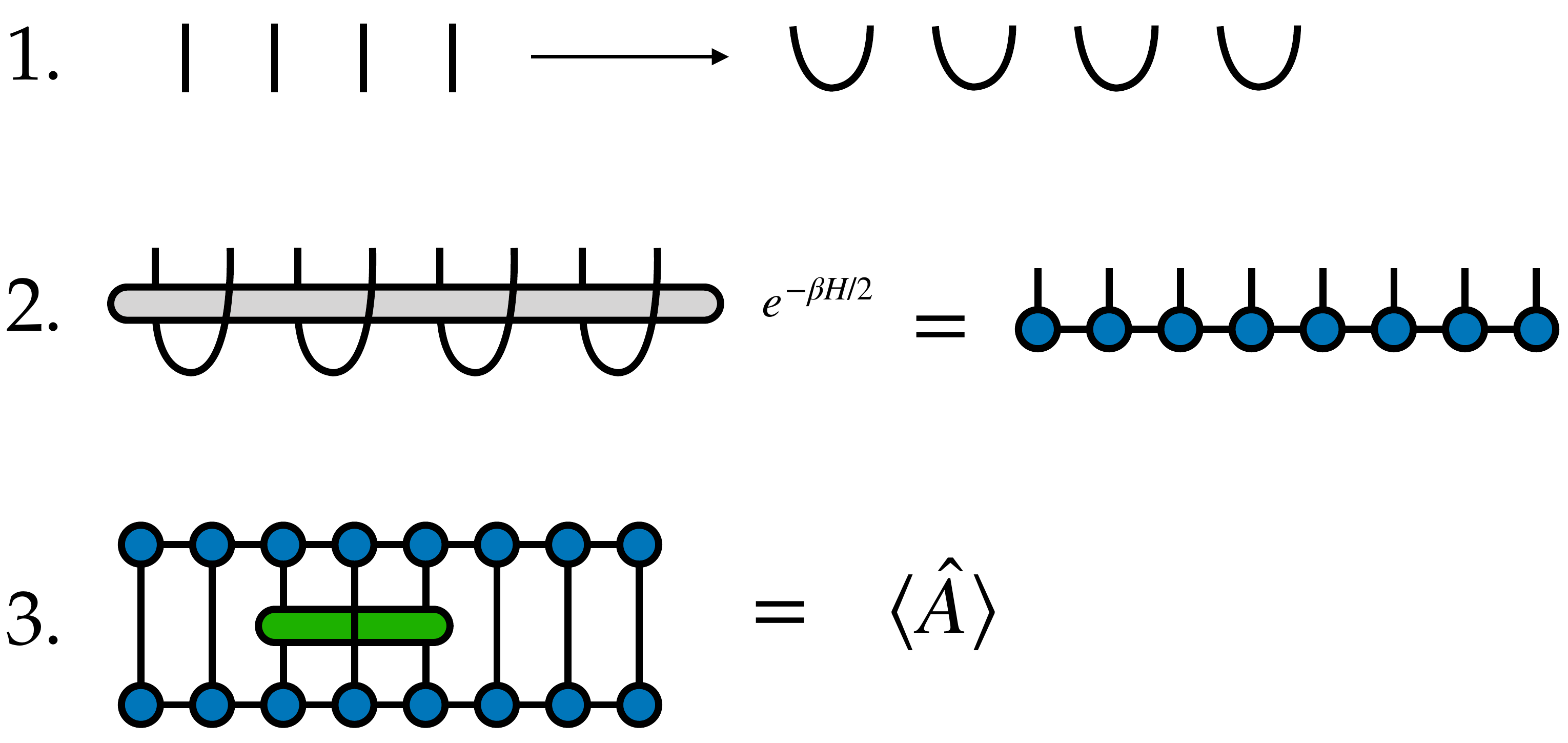}
\caption{Illustration of the ancilla method for obtaining a purified representation
of $e^{-H \beta/2}$. In step 1, the infinite temperature density matrix on $N$ sites is viewed as a 
pure state of $N$ maximally-entangled pairs. In step 2, the physical sites are evolved by imaginary
time $\beta/2$, and the state is scaled to have unit norm (not shown). Two copies of the evolved state are used in step 3 to compute the thermal average
of observables measured on the physical sites.
}
\label{fig:ancilla}
\end{figure} 

The goal of each method we will outline in this section is to obtain an estimate of the
thermal expectation value of local operators or observables $\hat{A}$, defined as
\begin{align}
\avg{\hat{A}} & = \frac{1}{\partZ} \Tr{\hat{A}\, e^{-\beta H}} \\
          & = \frac{1}{\partZ} \Tr{e^{-H \beta/2}\, \hat{A}\, e^{-H \beta/2}} \label{eqn:sqrtrho} \ .
\end{align}
The expression (\ref{eqn:sqrtrho}) above shows it is sufficient to compute $e^{-H\beta/2}$ to obtain
$\avg{\hat{A}}$, which is important for practical calculations because methods which compute $e^{-H \gamma}$ 
becomes increasingly expensive for larger $\gamma$ and because the symmetric form
of Eq.~(\ref{eqn:sqrtrho}) will be essential for the METTS and hybrid sampling techniques we
will describe.

Let us now briefly review two of the state-of-the-art methods for studying thermal systems
with tensor network methods: the ancilla method and the METTS algorithm. These two methods will be the building blocks for the ``hybrid'' method which is the main contribution 
of this paper.

\subsection{Ancilla Method}

The ancilla method starts from the observation that
\begin{align}
e^{-H \beta/2} & = e^{-H \beta/2}\ \mathbb{1}  \\
& = (e^{-H \tau})^M\  \mathbb{1}
\end{align}
where $\mathbb{1}$ is the identity operator on $N$ sites (in a Hilbert space ($\mathbb{C}^d)^{\otimes N}$), $d$ is the physical dimension on each site, for spin one Heisenberg model, $d=3$. 
The \emph{time step} is \mbox{$\tau = \beta/(2M)$} with $M$ (an integer) chosen large enough such
that $\tau \ll 1$.

To take advantage of existing tensor network methods for imaginary time evolving pure-state
wavefunctions, one makes the identification
\begin{align}
\mathbb{1} = \Bigg(\sum_{s=1}^d \ket{s} \bra{s}\Bigg)^{\otimes N} \rightarrow   \ket{\tilde{\mathbb{1}}}\stackrel{\text{def}}{=} \Bigg( \sum_{s=1}^d \ket{s}_{P} \ket{s}_{A} \Bigg)^{\otimes N}
\end{align}
where $\{\ket{s}\}_{s=1}^d$ is the orthonormal and complete computational basis on each site.
This identification amounts to viewing a many-body identity operator on $N$ sites as a 
pure-state wavefunction on $2N$ sites, where consecutive pairs of sites are in perfectly
entangled Bell pairs --- see Fig.~\ref{fig:ancilla}.1. The subscripts $P$ and $A$ above indicate that odd-numbered sites are viewed
as ``physical`` sites while even-numbered sites are fictitious ``ancilla'' sites whose role
is to thermalize the physical sites.

To compute thermal properties, one then carries out $M$ imaginary time steps $e^{-\tau H}$, acting
only on the physical sites. To carry out these steps, one can represent the state at each
imaginary time using a matrix product state (MPS) tensor network and time evolve using a variety of 
standard techniques, such as Trotter gates, the TDVP algorithm, 
or MPO techniques.\cite{Paeckel:2019,Bruognolo:2017} 
In this work we use the Trotter gate technique for its convenience and accuracy.
Taking expectation values of operators acting on the physical sites of the
final, time-evolved state is then equivalent to computing the thermal average Eq.~(\ref{eqn:sqrtrho}).
The key steps of the ancilla method are illustrated in Fig.~\ref{fig:ancilla}.


\begin{figure}[b] 
\includegraphics[width=0.85\linewidth]{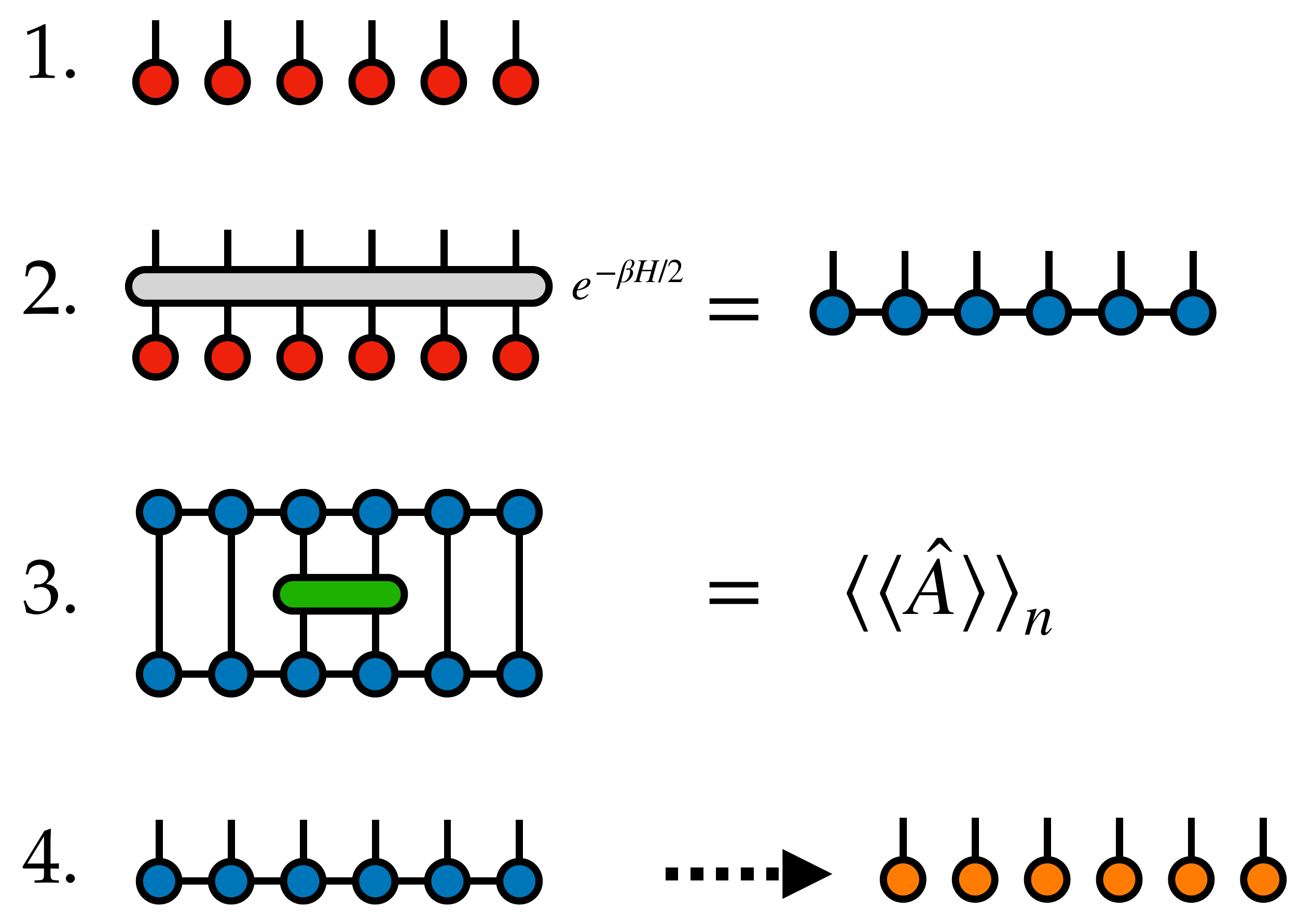}
\caption{Illustration of the METTS algorithm for obtaining 
the thermal expectation value $\avg{\hat{A}}_\beta = \Tr{\rho_\beta \hat{A}}$.
In step 1 the system is initialized to a product state. Step 2 consists of imaginary time evolving the product state
by a time $\beta/2$ and normalizing the resulting state. In step 3, one computes expectation values of physical observables $\hat{A}$
of interest, which serve as the Monte Carlo estimators of these observables. In step 4, one projectively measures each site, obtaining
a new initial product state which is used again in step 1.
}
\label{fig:metts}
\end{figure} 

\subsection{METTS Algorithm}

For minimal entangled typical states (METTS) algorithm, the thermal trace Eq.~(\ref{eqn:sqrtrho}) is taken  by a Monte Carlo sampling process.\cite{White:2009,Stoudenmire:2010,Bruognolo:2017,Binder:2017} Rather than computing the density matrix $e^{-H \beta/2}$, the physics is captured by a set of typical, low-entanglement pure states. 

For a given inverse temperature $\beta$, one 
starts from an arbitrary product state and evolves this state for $\beta/2$ imaginary time. The
resulting normalized state is the ``METTS'' wavefunction generated from the product state. After 
measuring physical expectation values of this METTS (which serve as the Monte Carlo 
estimators for the calculation results), 
the state is projectively measured or collapsed to a new product state, 
which becomes the initial state
for the next step. For efficiency, the collapse of the METTS into a new product state is performed by
performing a projective measurement on each site of the wavefunction sequentially. 
The steps of the METTS algorithm are illustrated in Fig.~\ref{fig:metts}. For a detailed introduction
to the METTS algorithm, see Ref.~\onlinecite{Stoudenmire:2010}.

\subsection{Hybrid METTS-Ancilla Algorithm}

The ancilla and METTS algorithms differ when dealing with the trace of density matrix occurring in the computation of physical observables. 
For the ancilla method, one traces out the ancilla degrees of freedom by an 
exact tensor contraction,
while for METTS, one uses a Monte Carlo approach to sample this trace over a product-state basis.
As we will argue, one can actually choose between either exact contraction or Monte Carlo sampling for each site separately.
Choosing exact contraction for a given site reduces the variance of observables involving that site, but at a possibly higher cost in terms of the entanglement
resulting from the imaginary time evolution.
We now describe a method that we call the hybrid METTS-ancilla method, or just 
\emph{hybrid method} for short, where some sites are traced by pairing with an ancilla and others are sampled by Monte Carlo. Like both METTS and ancilla, the hybrid method
remains unbiased and controlled with beneficial aspects of both the METTS and 
ancilla approaches.

The key question in developing the hybrid method is which physical degrees of freedom
should be paired with ancilla sites, and thus be traced exactly in a single step.
One motivation for choosing these sites is that degrees of freedom closer to
the spatial center of the system are typically more useful for estimating thermodynamic properties when using open boundary conditions,
as we do here. Therefore we will choose a contiguous ``cluster'' of sites at the spatial center to be paired with ancilla sites, leaving
all remaining ``environment'' sites to be sampled as in the METTS algorithm.
Though the resulting method resembles an embedding technique,
different choices of the cluster size does not bias the outcome in the limit of taking infinitely many samples.
This is because the methods to treat the cluster and environment are both unbiased, and
can be combined without introducing any uncontrolled approximations.
Because the method would obtain a converged result with
just a single sample if every site was paired with an ancilla, the intuition is that local
measurements well inside the cluster region should converge very quickly with number of samples.
We will see this is indeed the case.

\begin{figure}[b] 
\includegraphics[width=0.8\linewidth]{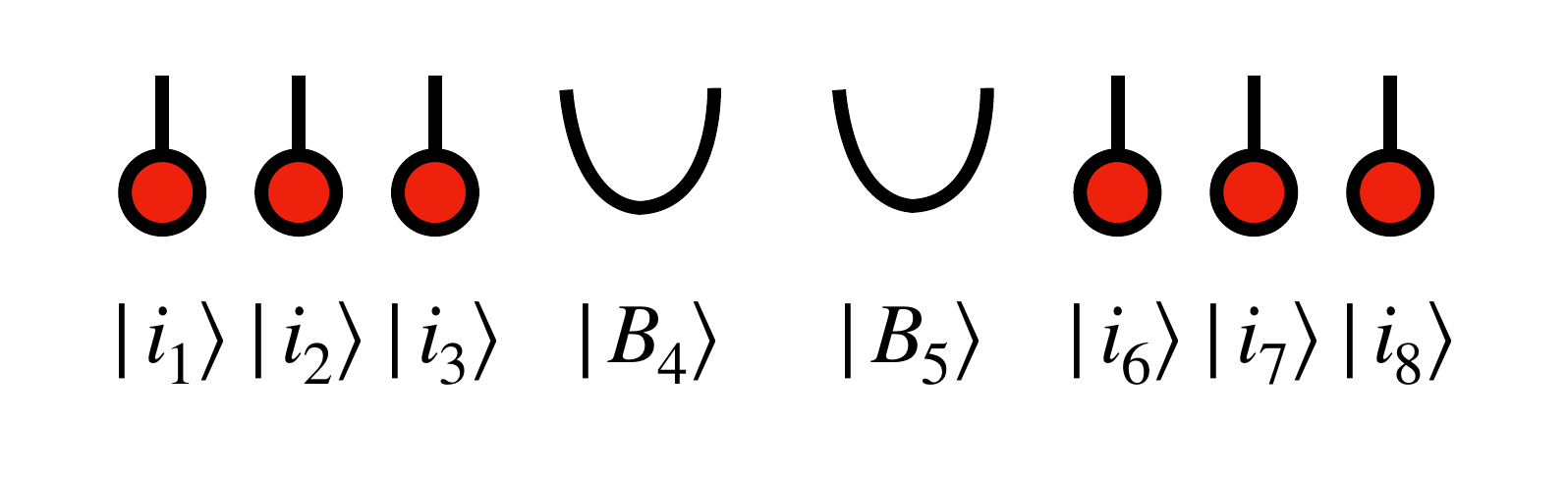}
\caption{The form the of initial state used in the hybrid METTS-ancilla method, with
sites 4 and 5 chosen to be the cluster sites and sites 1--3 and 6--8 the environment sites.}
\label{fig:initial}
\end{figure} 

For a one dimensional system such as shown in \Fig{initial}, we split the system into environment physical degrees of 
freedom (red dots in figure) and cluster physical degrees of freedom (left indices of arcs in figure). 
Here we take the cluster to have two physical sites for the purposes of illustration, but any bipartition of the sites into cluster and environment can be chosen. Throughout we will use indices $s_i$ to denote physical indices, whether of environment or cluster sites.

Similar to ancilla method, auxiliary degrees of freedom $a_j$ are introduced and
paired with each site of the cluster (right index of each arc). The system is prepared into an
initial state which is a product state of the environment sites, while the cluster is prepared
into a state consisting of products of maximally entangled Bell pairs:
\begin{align}
\ket{B_j} = \frac{1}{\sqrt{d}} \sum_{s_j, a_j=1}^d \delta_{s_j a_j} \ket{s_j} \ket{a_j} \ .
\end{align}

To begin deriving the hybrid ancilla-METTS algorithm, first note that expected values of physical observables $\hat{A}$ at finite temperature $T=1/\beta$ in the canonical ensemble can be written as
\begin{align}
\braket{\hat{A}} &=  \frac{1}{\partZ} \Tr{e^{-\beta \hat{H}} A} \\
& = \frac{1}{\partZ} \Tr{e^{-\beta \hat{H}/2} \hat{A} e^{-\beta \hat{H}/2}} \\
&= \frac{1}{\partZ} \sum_{i} \text{Tr}_{C} \bra{i}_E  e^{-\beta \hat{H}/2}  \hat{A} e^{-\beta \hat{H}/2} \ket{i}_E   \\
&= \frac{1}{\partZ} \sum_{i} P_i \bra{\psi_i}  \hat{A} \ket{\psi_i} \ . \label{eq:eq2}
\end{align}
Here the sum over $i$ denotes a sum over all product states $\ket{i}_E$ on the environment sites and $\text{Tr}_C$ denotes
a trace over the physical cluster sites.

The above form motivates an algorithm where one samples the pure states $\ket{\psi_i}$ with probability $P_i/\partZ$
where the states $\ket{\psi_i}$ are defined as
\begin{align}
\ket{\psi_i} & = \frac{1}{\sqrt{P_i}} e^{-\beta \hat{H}/2} \ket{i}_E \ket{B}_C \label{eqn:psi} \\
\ket{B}_C & = \prod_{j\in C} \ket{B_j}
\end{align}
with Bell-pair states $\ket{B_j}$ defined as above and the factor $P_i$ is defined such that the $\ket{\psi_i}$ are normalized,
thus
\begin{equation}
P_i = d^{N_C} \bra{i}_E \text{Tr}_C \big[e^{-\beta \hat{H}}\big] \ket{i}_E \ .
\end{equation}
where $N_C$ is the number of physical sites in the cluster region.
In practice the $P_i$ are not explicitly computed, but arise implicitly through the preservation of 
the unit norm of each pure state throughout the imaginary time evolution used to compute the $\ket{\psi_i}$.

\subsubsection{Markov Chain Sampling Algorithm}
 
Similar to METTS method, we generate the $\ket{\psi_i}$ through a Markov sampling process. 
The steps of this sampling algorithm are illustrated in Fig.~\ref{fig:illustration}
and below we will show that the algorithm satisfies detailed balance with respect to the probability weights $P_i$. 
The algorithm proceeds as follows:
\begin{enumerate}
\item Begin step $n$ from an initial state $\ket{i_n}_E \ket{B}_C$ which is a product state $\ket{i_n}_E$ over the environment sites
and a product of physical-ancilla Bell pairs for the cluster sites.
\item Evolve this state by an imaginary time $\beta/2$ to obtain $\ket{\psi_{i_n}}$
as defined in Eq.~(\ref{eqn:psi}) above. For this step we use an MPS representation of the entangled $\ket{\psi_{i_n}}$ state and 
the Trotter-gate time evolution approach, though other representations and algorithms could be used.
\item Compute the Monte Carlo estimator of each physical observable $\hat{A}$ as $\bra{\psi_{i_n}}\hat{A}\ket{\psi_{i_n}} \stackrel{\text{def}}{=} \langle\langle \hat{A} \rangle\rangle_n$.
\item Collapse the environment sites to a new product state $\ket{i_{n+1}}_E$ via projective measurements of each environment site, then re-initialize the
cluster region to perfect Bell-pair states for each physical-ancilla site pair within the cluster region.
\end{enumerate}

%
%


\subsubsection{Discussion of Detailed Balance}

\bfg
\centering
\includegraphics[width=1.0\linewidth]{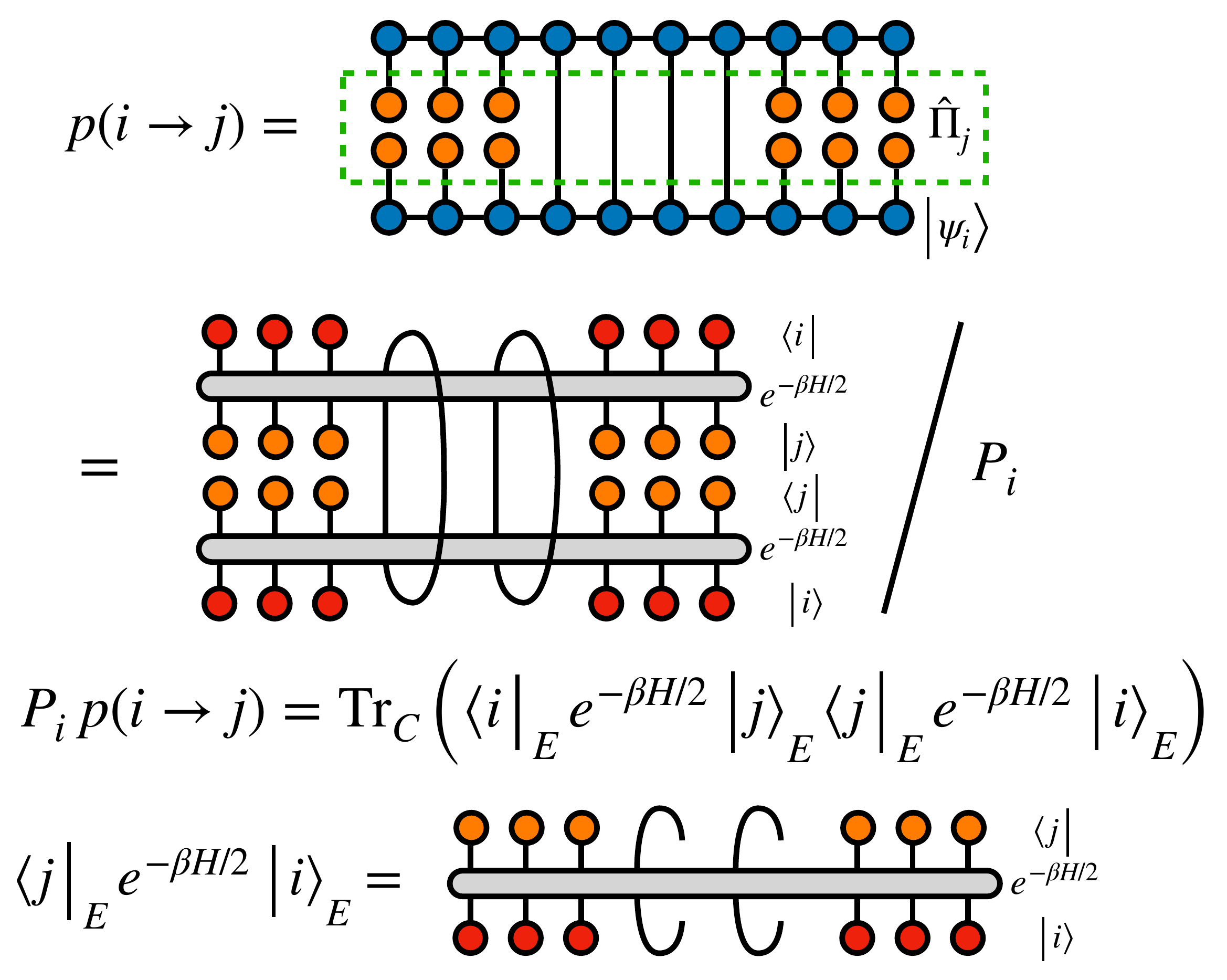}
\caption{Illustration of steps to show the preservation of detailed balance \Eq{detailbalance1} within the hybrid METTS-ancilla algorithm. 
The red and orange dots represent the initial product states $\ket{i}_E$ and $\ket{j}_E$, respectively.  }
\label{fig:detailbalance}
\efg

Let us define $p(i \rightarrow j)$ as probability that the collapse procedure discussed above generates a new initial state 
$\ket{j} = \ket{j}_E \ket{B}_C$ from a previous entangled state $\ket{\psi_{i}}$.
This transition probability can be written as the expectation value of a 
projector $\hat{\Pi}_{j} = \left( \ket{j}_E\bra{j}_E \right) \otimes \mathbb{1}_{C}$, where $\mathbb{1}_{C}$ is an identity operator 
acting within the cluster (both physical and ancilla sites). Using this projector, we can write out the transition probability as
\beqa
p\big( i \rightarrow j \big) & = & \braket{ \psi_i | \hat{\Pi}_{j} | \psi_i } \nonumber \\
					  			& = & \frac{1}{P_i} \ \text{Tr}_C\!\! \left[  \bra{i}_E e^{-\beta H/2 }\ket{j}_E  \bra{j}_E e^{-\beta H/2 } \ket{i}_E \right]  \nonumber \\
\ \ \label{eq:detailbalance1}
\eeqa
where we have used the definition of the state $\ket{\psi_i}$ from Eq.~(\ref{eqn:psi}). Note that $\bra{i}_E e^{-\beta H/2 } \ket{j}_E$ 
and its transpose are operators in the Hilbert space of the cluster, which is why a further trace over the cluster sites 
is needed above to obtain a scalar quantity; see \Fig{detailbalance} for an illustration.

From the cyclic property of the trace operation, one can see by inspection of Eq.~(\ref{eq:detailbalance1}) that 
$P_i\cdot p(i\rightarrow j)$ is a symmetric function of $i$ and $j$. Thus it immediately follows that
\beq
 \frac{P_i}{\partZ} \, p\big( i \rightarrow j \big) = \frac{P_j}{\partZ}\, p\big( j \rightarrow i \big)
\eeq
and thus detailed balance is respected by the hybrid METTS-ancilla sampling algorithm.

\section{Results} 

For an example system to test and benchmark the hybrid method described above, we will 
consider the nearest-neighbor $S=1$ Heisenberg antiferromagnet on a two-leg ladder with
open boundary conditions in both directions. The Hamiltonian is written as
\beq
H = J \sum_{i=1}^{L-1}\left(\hat{S}_{i,1} \cdot \hat{S}_{i+1,1}+\hat{S}_{i,2} \cdot \hat{S}_{i+1,2}\right)+J_{\perp} \sum_{i=1}^{L}\hat{S}_{i,1} \cdot \hat{S}_{i,2}
\eeq
where $J >0$ is the interaction strength within the chain and $J_{\perp}>0$ is the coupling strength between the two chains; $L$ is the length of ladder. In our calculation we set $J_{\perp}=0.1\,J$, because we want the system to have relatively large entanglement
so as to resemble two-dimensional systems or more challenging classes of Hamiltonians where we anticipate the hybrid method will be more advantageous. Note that in the $J_{\perp}=0$ limit, the system decouples into two chains and so the  entanglement is twice that of a single chain, resulting in squared
 MPS bond dimension compared to a single chain.

Our calculations have three main sources of error, which are all controlled. First, the Trotter-Suzuki error, because we split the imaginary time evolution into many small steps. This error is controlled by the choice of the time step $\tau$, which we choose to be 
$\tau = 0.05$, and by using a second order Trotter-Suzuki decomposition, so that the error scales as $O(\tau^3)$ per time step. Secondly, all the physical quantities are averaged over different Monte Carlo samples, resulting in a statistical variance shown by 
an error bar in our results. In the limit of many samples $N$, the sampling error is proportional to $1/\sqrt{N}$. The sampling error is easily controlled by taking a larger number of samples. Thirdly, there is truncation or cutoff error. 
Carrying out a time evolution step on each bond locally
destroys the MPS form. To recover this form, one performs a singular value decomposition, which must be truncated to control the overall costs of the algorithm.
The resulting truncation error can be calculated by summing the squares of the discarded singular values (or Schmidt weights), divided by the sum of squares of all singular values. 
In practice we set this truncation error to a fixed target value, which automatically
determines the maximum number of Schmidt weights that can be discarded while still giving an accurate result.

\subsection{Computing Physical Observables}

To compare the ancilla, METTS, and hybrid methods, we measure the energy and the uniform magnetic susceptibility at inverse temperatures
\mbox{$\beta = 4, 8$}. To make the comparisons fair, 
we choose an different way of estimating observables for each method, with the goal of 
giving each method the best chance to converge as quickly as possible.
For the hybrid method, we  select the middle $16\times2$ physical sites to be the cluster sites with ancilla pairs, then measure observables only on the middle $6\times 2$ physical sites, since the sites at the edge of the cluster have more statistical variance due to influence from the sampled sites nearby.
In contrast, for the METTS algorithm the best approach is 
to measure observables over most of the system, since 
statistical fluctuations are reduced by averaging over many sites. So we primarily
measure the center $80\times2$ bulk sites, though we also present results for the
center $6\times2$ sites to emphasize the different amount of fluctuations compared to 
the hybrid method (which measurement approach was used for METTS 
is labeled in the legends of
each figure).
Finally, for the ancilla method we measure the central $6\times2$ sites since there are
no statistical fluctuations and the main consideration is reducing finite-size effects
due to the open boundaries.

\bfg
\centering
\includegraphics[width=.95\linewidth]{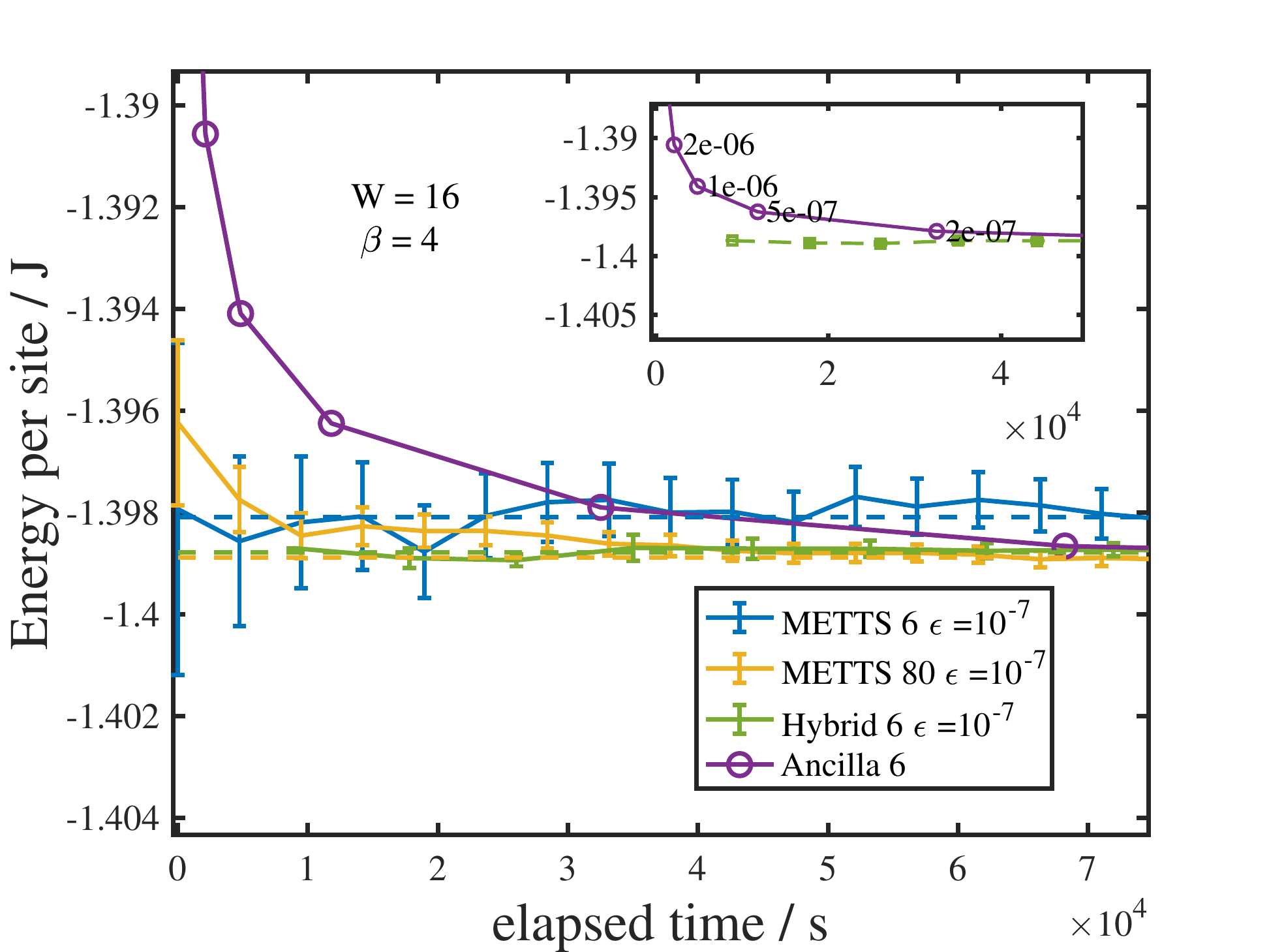}
\caption{ Hybrid method energy convergence compared with METTS and ancilla method at $\beta = 4$ for the $100\times2$ Heisenberg ladder. The length of the cluster used for the hybrid method is $W=16$, meaning the middle $16\times2$ sites.
The $x$ axis is the time spent in seconds. The $y$ axis is the energy per length. 
The truncation error cutoff for the hybrid and METTS algorithms used was $10^{-7}$. 
Error bars indicate the Monte Carlo sampling error for the METTS and hybrid algorithms.
Ancilla results are shown for various truncation error cutoffs indicated next
to each point. The orange curve is the averaged energy over the  bulk of $80\times2$ sites. The other curves show measurements over the center $6\times2$ physical sites. The dashed line donates the converged value with the same color. In the upper right pane, we zoom in and show some details, and display every sample computed by the hybrid method. }
\label{fig:ladderb4E}
\efg

Figure~\ref{fig:ladderb4E} shows the energy per site of the system 
and Fig.~\ref{fig:ladderb4X} the uniform susceptibility per site at \mbox{$\beta=4$}
estimated by each method versus the elapsed CPU time spent in seconds.
To simplify the comparison, we use only one CPU core for each case, although 
it is important to note that the hybrid and METTS methods could be straightforwardly parallelized. Dashed lines correspond to the converged energy of each method shown with the same color, and were obtained by running the method for 
longer times than are shown in the figure. For the hybrid and METTS algorithms, we used a truncation error cutoff of $\epsilon=10^{-7}$, and computed
many hundreds of samples. In order simplify the figure, we show only selected data points
for a fixed time period versus every sample.  The ancilla method does not involve sampling:  for a fixed truncation cutoff $\epsilon$, it takes a particular time to complete and
then the energy is obtained. The data points shown for the ancilla are results 
obtained with different $\epsilon$ ranging from $10^{-5}$ to $10^{-7}$, 
from which one can estimate the error relative to the exact result.

For the energy computations shown in Fig.~\ref{fig:ladderb4E}, we can see that
the hybrid method converges about six to eight times as quickly as a similarly converged ancilla calculation ($\epsilon=10^{-7}$), 
relative to the value obtained when running each method for much longer times. 
The hybrid method is even more advantageous when compared to METTS averaged over the central $80\times2$ sites.

\bfg
\centering
\includegraphics[width=1\linewidth]{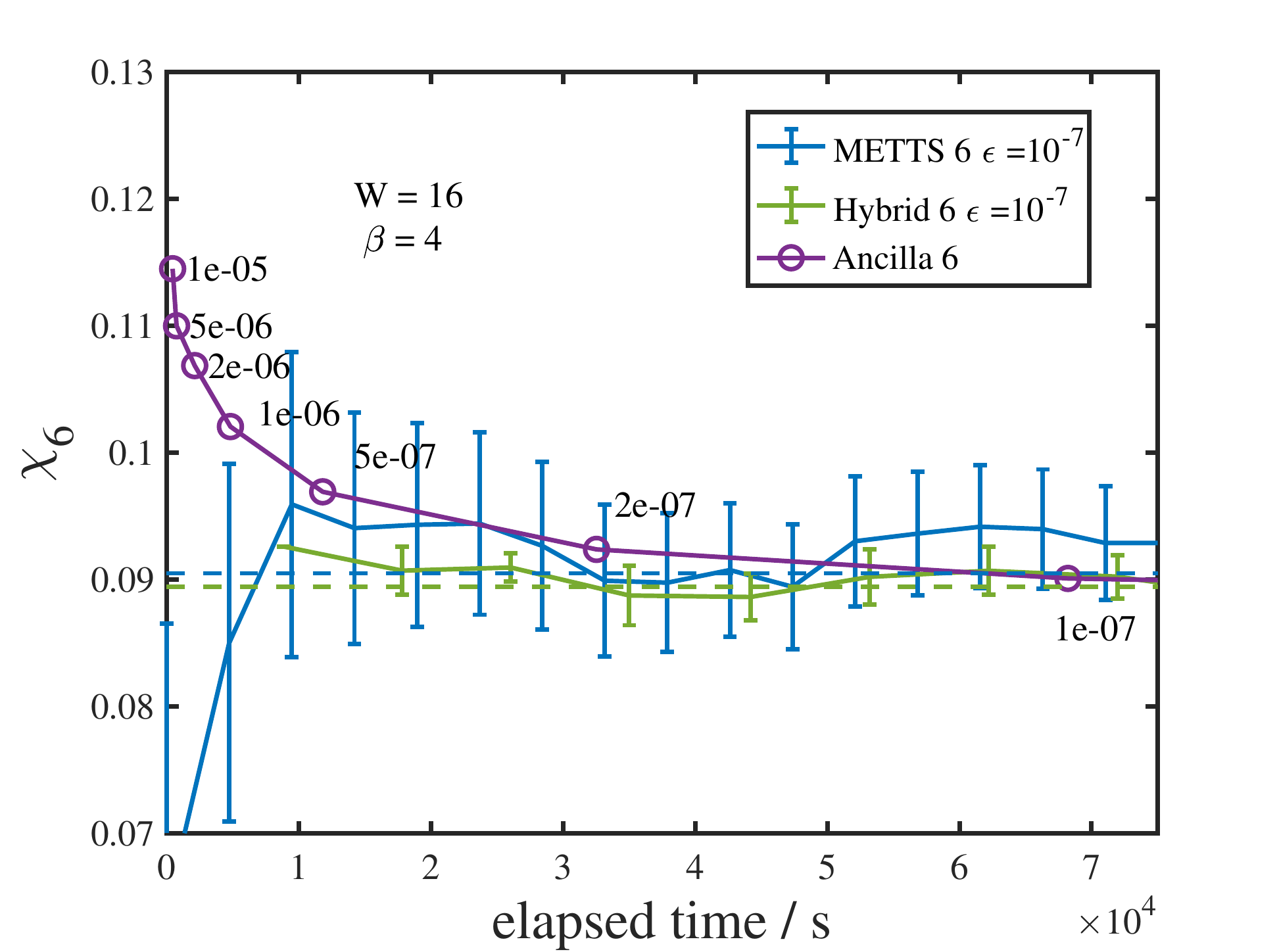}
\caption{ Susceptibility convergence of different methods at $\beta=4$ of $100\times2$ Heisenberg ladder. The susceptibility are calculated by \Eq{sus}. The x axis denote the time cost in seconds. The sampling error bar of METTS and Hybrid method are given. The cutoff and meaning of legends are the same as \Fig{ladderb4E}. }
\label{fig:ladderb4X}
\efg

We also compare susceptibility in \Fig{ladderb4X}.  The meanings of symbols and colors are the same as \Fig{ladderb4E}.  The susceptibility is an extensive value and the bulk is close to translational invariant, so the susceptibility for a non-translation-invariant 
system can be estimated by 
\beq
\chi=\frac{\beta}{3} \sum_i{\avg{\vec{S}_{c}\cdot\vec{S}_{i}}}
\eeq 
where $c$ is a site at the center of the system. 
To obtain more accuracy, we average over the middle $6\times2$ sites as
\beqa
\chi_6\, \stackrel{\text{def}}{=}\  \frac{\beta}{3}\frac{1}{12}\sum_{j \in 6\times2}\sum_{i \in \text{all}}
\avg{\vec{S_j}\cdot\vec{S}_{i}}, 
\label{eq:sus}
\eeqa
where $i$ runs over all the spins. The error bar of $\chi$ from hybrid method is smaller compared to METTS method.  Results from all three methods agrees with each other.  
For the hybrid method, the error bars for the susceptibility are larger compared with the energy in \Fig{ladderb4E}. There are two reasons. 1. The magnitude of $\chi$ is in the order of $0.1$, which makes the relative error magnified. 2. From the definition of $\chi$ in \Eq{sus},  measurement of $\chi$ necessarily involves spins outside of the cluster.

\bfg
\centering
\includegraphics[width=1\linewidth]{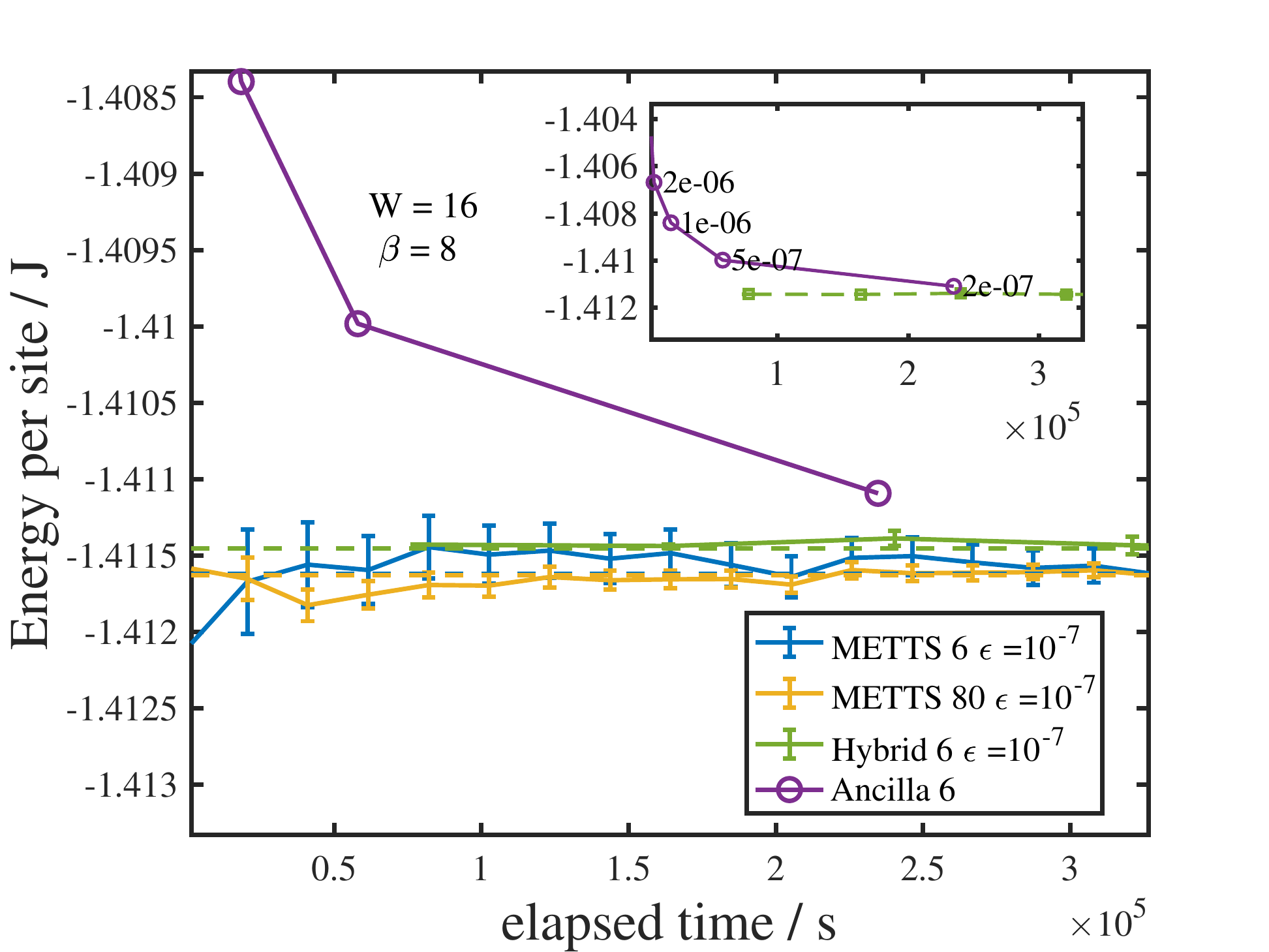}
\caption{ Hybrid method energy convergence compared with METTS and ancilla method at $\beta = 8$ of $100\times2$ Heisenberg ladder.  The cutoff and meaning of legends are the same as \Fig{ladderb4E}. We take only four samples by hybrid method within the time scope.  The cutoff $\epsilon$ of ancilla ranges from $10^{-5}$ to $2\times10^{-7}$. }
\label{fig:ladderb8E}
\efg

Calculations for $\beta=8$ take roughly one order of magnitude longer than for
$\beta=4$ when using the same cutoff error within each method. 
This is due to longer imaginary time evolutions. The entanglement of each 
pure state becomes larger and results in much larger MPS bond dimensions. 
The ancilla suffers the most from this effect, so that we can no longer obtain a 
result for cutoff  \mbox{$\epsilon=10^{-7}$} within one week. 
The METTS is the least affected. 
For $\beta=8$, we choose the size of the hybrid method cluster to be $W=16$ to 
balance the entanglement growth and sampling variance. 
We again choose a cutoff $\epsilon=10^{-7}$ for the hybrid and METTS methods, and cutoffs of $10^{-5}$ to $2\times10^{-7}$ for the ancilla method.  
From Figs.~\ref{fig:ladderb8E} and \ref{fig:ladderb8X}, one can see the error bars of the sampling methods are smaller compared to \mbox{$\beta=4$}, since at low temperature each sampled wavefunction $\ket{\psi_i}$ 
becomes very close to the ground state.

\bfg
\centering
\includegraphics[width=1\linewidth]{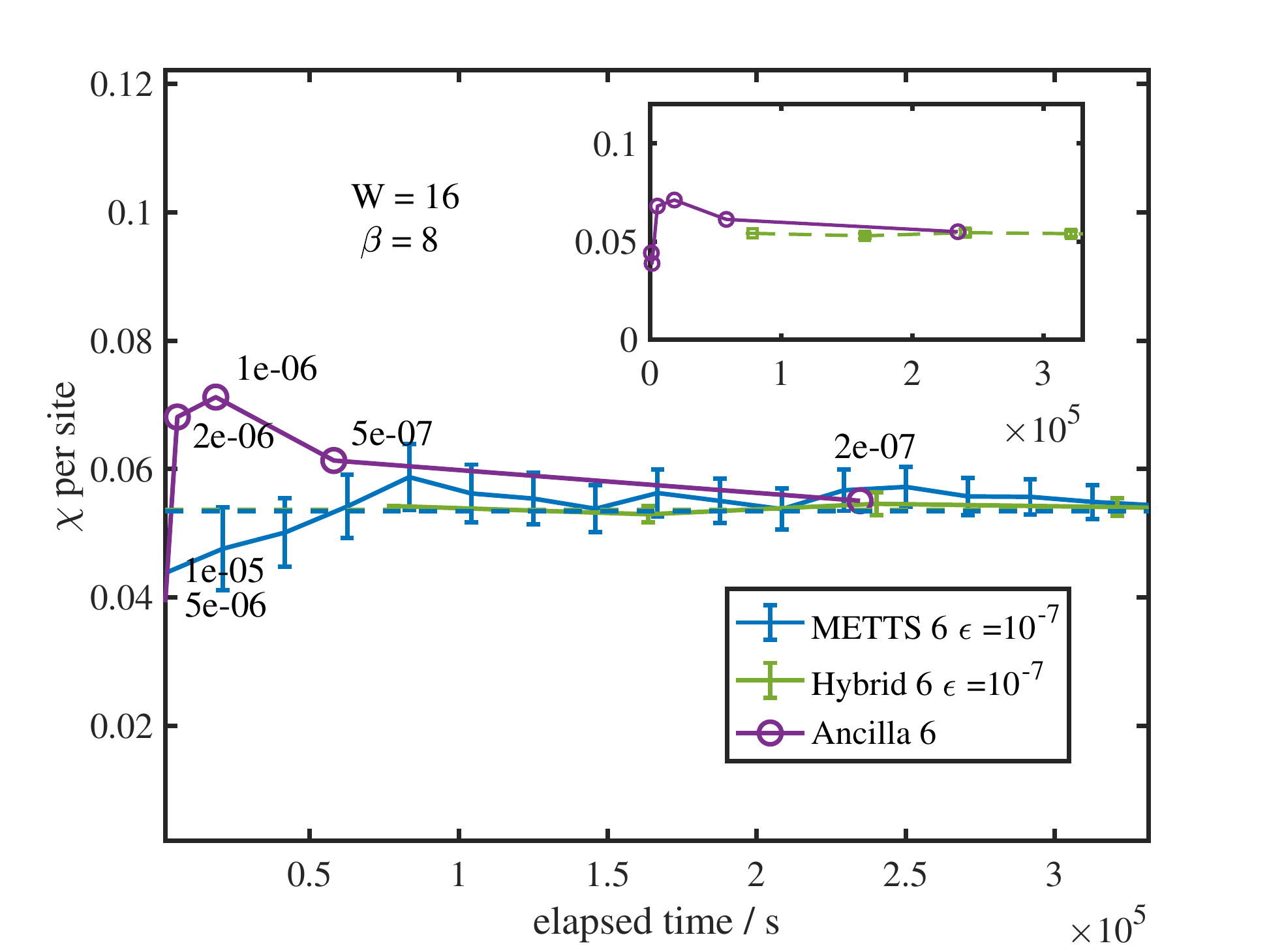}
\caption{ The susceptiblity of $100\times 2$ ladder at $\beta=8$. The meaning of 
the legend is the same as \Fig{ladderb4X}. All hybrid samples are shown in this figure. }
\label{fig:ladderb8X}
\efg

\subsection{Entanglement and Cost of Hybrid States}

The hybrid method strikes a balance between the cost of representing each sample
and number of samples necessary, which can be adjusted by the choice of the cluster size.  When the cluster covers the entire system, only one sample is needed and the method becomes identical to the ancilla method, which is the most efficient method at high temperatures.
In the other limit of no cluster sites, the hybrid method becomes identical to the METTS algorithm. METTS has the benefit of dealing with lower-entanglement states and thus becomes the best method at very low temperatures. 
For intermediate temperatures, taking a modest cluster size in the hybrid method 
such as $W=16$ significantly reduces the sampling error without too much increase in the entanglement of  each sample.

\bfg
\centering
\includegraphics[width=1 \linewidth ]{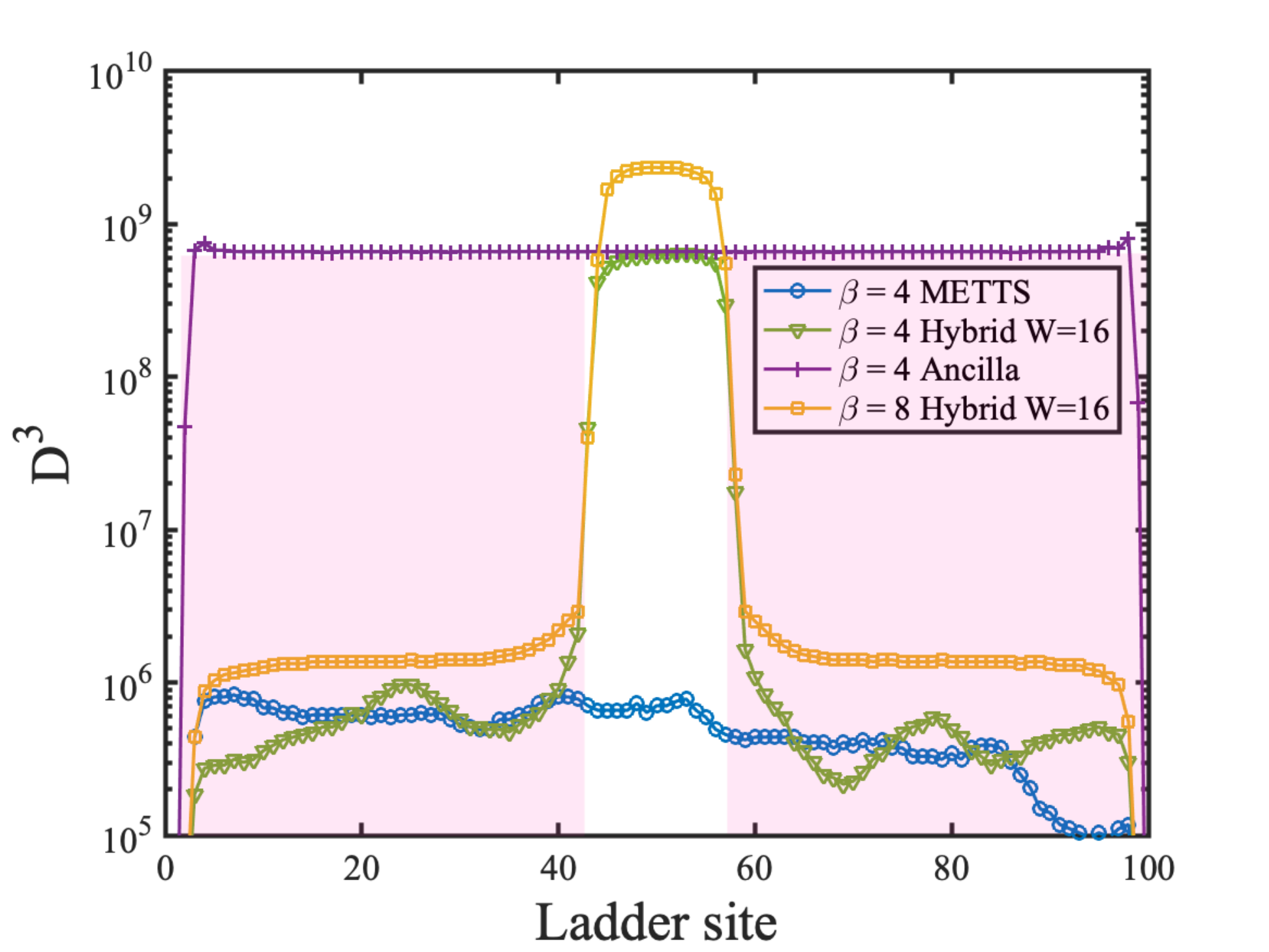}
\caption{ The spatial distribution of bond dimension of the MPS for the ladder at $\beta=4$ and $8$ with truncation error cutoff $\epsilon=10^{-7}$. The time cost of making one MPS wavefunction within each method can be approximated by $D^3$ integrated over time, which is the area under the curve. Thus the computational savings of making one hybrid-method state versus the ancilla state can be visualized by the area of the pink shadow. Note that the vertical axis uses a log scale; thus the visible fluctuations in the bond dimension of the hybrid and METTS samples chosen for $\beta=4$ do not represent a very significant effect when comparing the time to make one of these states versus the ancilla-method state.
} \label{fig:D}
\efg

To visualize the computational effort required in each method, in \Fig{D} we plot the cube of the typical MPS bond dimension $D^3$ along the ladder for a representative pure-state
sample within each method.
Outside of the cluster region, the bond dimension $D$ of a hybrid sample is similar to
a METTS sample, and much smaller compared with the state computed by the ancilla method.
Inside of the cluster region, the bond dimension of the hybrid method is very similar
to the ancilla state computed with the same truncation cutoff.
The computational cost of producing each sample scales as $O(D^3)$, 
so the area under the curve in \Fig{D} shows the cost of producing one sample 
in each method. Observe that most of the computational cost of the hybrid method comes from the cluster in the center. 
So the hybrid method saves large amount of work by a factor of $\frac{W}{L N_s}$, 
where $W$ is the length of the cluster, $L$ the length of the whole system, and $N_s$
the number of hybrid samples needed. For low to intermediate temperatures, as few as just
$N_s=2$ hybrid samples can be all that is needed to reach convergence.
In the figure, we also show a hybrid state at a lower temperature $\beta=8$ to illustrate
the growth of computational effort with decreasing temperature that is inherent to both
the hybrid and ancilla approaches.

\bfg
\centering
\includegraphics[width=1 \linewidth ]{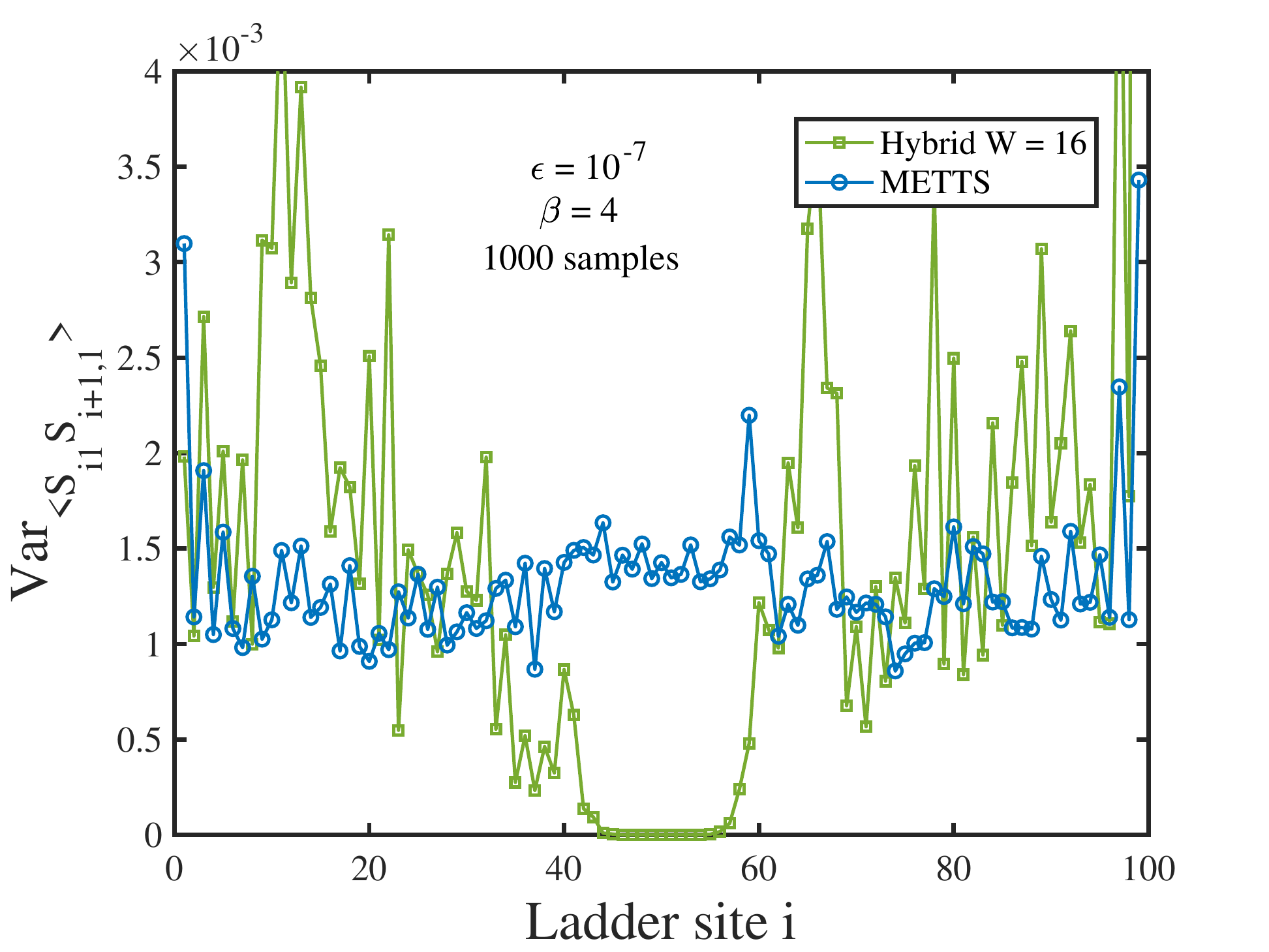}
\caption{The variance value of nearest neighboring energy along the ladder $\braket{ \vec{S}_{i1} \cdot \vec{S}_{i+1,1}}$.
The variance of both the hybrid and METTS methods at $\beta=4$ are estimated from 1000 samples.  
} \label{fig:Var}
\efg

In \Fig{Var} we plot the variance of the energy of each bond $\braket{ \vec{S}_{i,1}\cdot \vec{S}_{i+1,1}}$ along the first leg of the ladder. The results of both the hybrid and METTS methods are shown for $\beta=4$. 
For the METTS method, the variance along the chain is very similar for every bond 
throughout the bulk of the system, and somewhat higher close to each end.
In contrast, the variance of the hybrid method is greatly reduced in the cluster region,
supporting our finding that it is advantageous to estimate observables using only
cluster sites as much as possible.

\bfg
\centering
\includegraphics[width=1 \linewidth ]{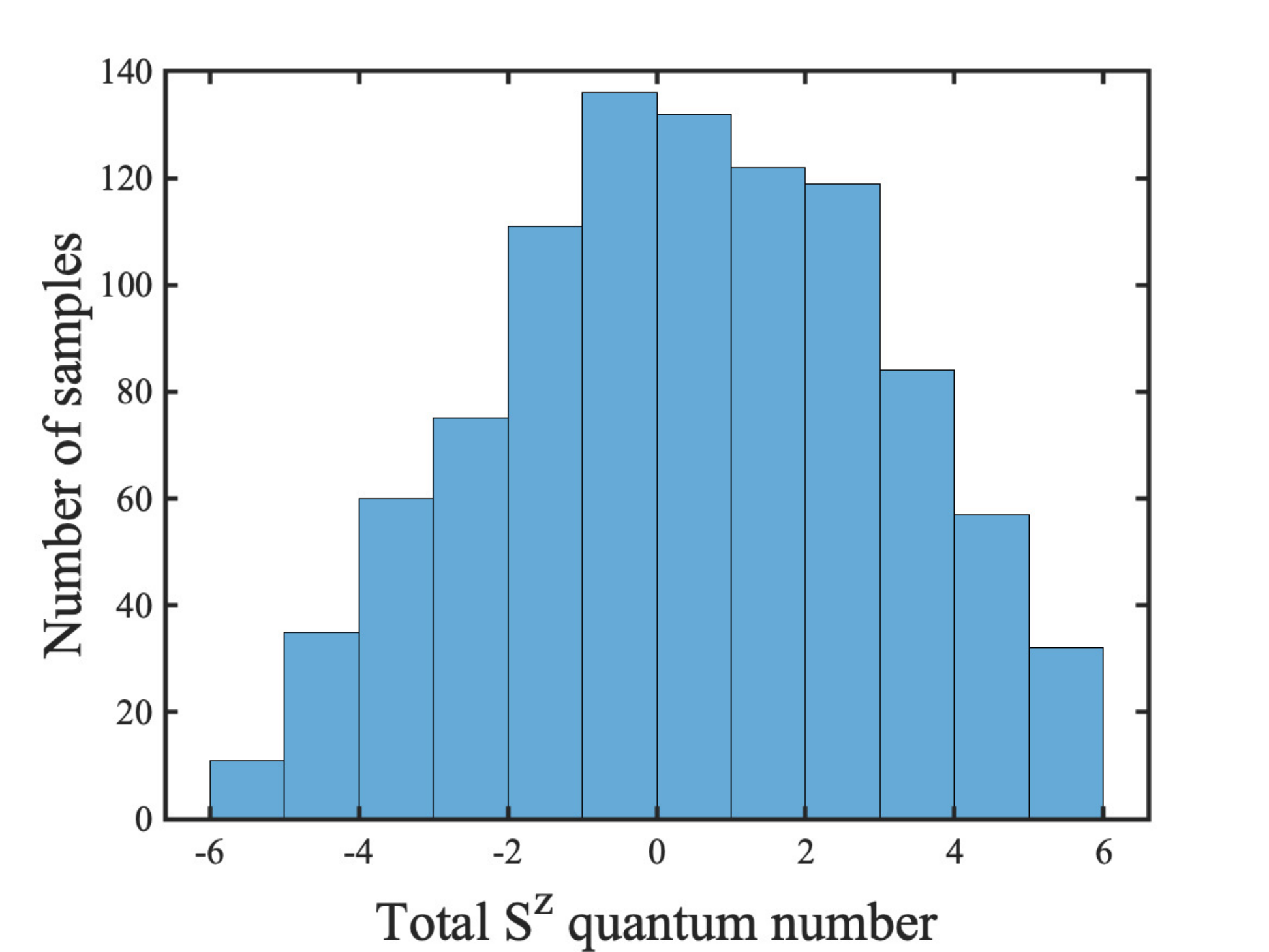}
\caption{ The histogram of the total $S^z$ quantum number of the hybrid method with $16\times2$ cluster sites on a $100\times2$ ladder at temperature \mbox{$\beta=4$}. The vertical axis shows number of samples observed to have the same quantum number. 
The process of partially collapsing the previous sample to produce a new initial state
in the hybrid method allows the next initial state to have different total quantum
numbers from the previous one. \label{fig:qn}}
\efg

\subsection{Exploiting Conserved Quantum Numbers}

A technical yet important advantage of the hybrid method over the METTS method 
is that the hybrid method automatically allows the total quantum numbers to
fluctuate from sample to sample, making it
possible to sample in the grand canonical ensemble while still realizing the benefits
of conserved quantum numbers within each imaginary time evolution step.
Recall that conserving quantum numbers when time evolving an MPS allows the MPS
tensors to have a block-sparse structure, which can significantly speed up  
computations.\footnote{The magnitude of the speedup due to quantum number block
sparsity depends on system-specific 
details, such as the number of conserved quantities and the quantum number 
fluctuations at a given temperature.}
However, the METTS algorithm suffers from a technical drawback where collapsing 
all of the sites to produce the next initial state conserves 
the total quantum numbers. Thus METTS remain stuck in a single quantum
number sector. Though tricks exist to fix this problem in special settings, 
such as for SU(2) invariant spin models,\cite{Stoudenmire:2010} in general the 
only other solution known involves time evolving some fraction of the METTS 
without using quantum numbers.\cite{Binder:2017}

In the hybrid method, however, only the environment sites are collapsed after each
time evolution step. The total quantum number of the collapsed environment
sites can vary depending on the outcomes of the projective measurements for the 
simple reason that the environment sites are only a subset of the whole system. 
After the environment collapse is done, but before restoring the physical-ancilla Bell pairs
within the cluster region, the total quantum number of the cluster region 
also varies such that the quantum number of the entire system remains conserved.
Thus when the state of the cluster is discarded and replaced with perfect Bell pairs,
which can always be defined with flipped quantum numbers on the ancilla sites such
that the restored cluster has total quantum number zero, the new initial state will 
have a total quantum number equal to that of just the environment sites after the collapse.

In Fig.~\ref{fig:qn} we demonstrate empirically that the quantum numbers of hybrid method
samples do indeed fluctuate. A histogram of the total-$S^z$ quantum number of samples
computed for a Heisenberg ladder at temperature $\beta=4$ shows significant fluctuations 
peaked around a total $S^z$ of zero. However, during the time evolution step used
to produce each sample, the total quantum number does not fluctuate and thus this 
costly step can benefit from block-sparse tensors.

\section{Conclusion and outlook}

We have proposed a method to hybridize two methods for simulating finite-temperature 
quantum systems with tensor networks, one based on purification (ancilla method) and 
the other based Monte Carlo sampling (METTS method). The resulting algorithm resembles an embedding method, with a cluster of system sites having ancilla partners, and the 
remaining environment sites sampled using Monte Carlo. 
By paying the price of more entanglement versus METTS but much less entanglement 
versus ancilla, the hybrid method gains a large reduction in the number of samples 
needed to converge properties measured within the cluster. We thus find the hybrid method 
can be superior to both the ancilla and METTS for a wide range of intermediate temperatures.
The hybrid method also solves an important technical issue with the METTS approach that prevents METTS from taking full advantage of quantum number conservation.\cite{Binder:2017}

It is important to note that all calculations here were performed on one CPU. 
By computing samples in parallel, both the hybrid and METTS method 
can be converged much more quickly as no communication is required across the parallel
computers.
Given the advantage of the hybrid method over the ancilla method even on a single CPU, 
the possibility of parallelizing it should make it that much more advantageous.

When treating two dimensional systems with MPS tensor networks, 
the bond dimension needed grows very quickly with system size, often reaching many 
thousands for ladders of transverse size of order ten. In this setting, we also expect 
the advantages of the hybrid method to stand out even more, since the sensitivity of the 
MPS-ancilla method to entanglement makes it quite costly in two dimensions.\cite{Bruognolo:2017}

In this paper, we chose a cluster of sites in the center of the system to be 
paired with ancilla sites.  But it is straightforward to implement the hybrid method 
for other choices of which sites are paired. For example, it would be interesting to 
try interleaving paired and unpaired sites along a chain to see if there is a 
computational advantage. One can also envision dynamically collapsing ancilla
sites during the time evolution based on some physical criterion, such as when 
ancilla sites become  nearly disentangled from the rest of the system. 

Finally, techniques for evolving projected entangled pair state (PEPS) 
2D tensor networks have been developed and used successfully within
the ancilla approach to study challenging systems such as the Hubbard 
model.\cite{Czarnik:2012,Czarnik:2015,Czarnik:2016,Czarnik:2019,Czarnik:2019a}
A promising direction would be to develop the hybrid METTS and ancilla
approach for PEPS tensor networks, which might mitigate the otherwise high costs of 
using them.

\begin{acknowledgments}
We thank Steven R. White for many helpful discussions on the possibility of 
combining the ancilla and METTS algorithms. The Flatiron Institute is a division 
of the Simons Foundation.
\end{acknowledgments}

\bibliography{hybrid}

\end{document}